\documentclass[11pt,a4paper]{article}
\usepackage[a4paper,
left=2.5cm,
right=2.5cm,
top=2.5cm,
bottom=2.5cm]{geometry}

\usepackage{lmodern}
\usepackage[colorlinks,hypertexnames=false]{hyperref}
\usepackage{caption}
\usepackage{cite}

\usepackage{amsmath}
\usepackage{amssymb}
\usepackage{amsfonts}
\usepackage{amsthm}
\usepackage{bm}
\usepackage{mathrsfs}

\usepackage{graphicx}
\usepackage{booktabs}
\usepackage{multirow}
\usepackage{array}

\usepackage[numbers,sort&compress]{natbib}

\usepackage{hyperref}

\begin{document}

\begin{center}

{\LARGE\bfseries
Effects of Curvature--Scalar Coupling on Vacuum Energy in Flat
(3+1)-Dimensional Space-Time
\par}

\title{}
\author{}
\date{}

\vspace{1cm}

{\large
Volodymyr Gorkavenko$^{1,2}$,
Oleh Barabash$^{1}$,
Pavlo Nakaznyi$^{3}$,
Mariia Tsarenkova$^{1}$,\\
Nazar Yakovenko$^{1}$,
Andrii Zaporozhchenko$^{1}$
}

\vspace{8mm}

{\it
$^{1}$Faculty of Physics,
Taras Shevchenko National University of Kyiv,\\
64/13 Volodymyrska Street,
Kyiv 01601,
Ukraine
}

\vspace{3mm}

{\it
$^{2}$Bogolyubov Institute for Theoretical Physics,
National Academy of Sciences of Ukraine,\\
14-b Metrolohichna Street,
Kyiv 03143,
Ukraine
}

\vspace{3mm}

{\it
$^{3}$Institute of Physics and Technology,
Igor Sikorsky Kyiv Polytechnic Institute,\\
37 Beresteiskyi Avenue,
Kyiv 03056,
Ukraine
}

\vspace{5mm}

{\small
}

\end{center}


\begin{abstract}
We investigated how a magnetic topological defect affects the vacuum polarization of a charged massive scalar field in a flat $(3+1)$-dimensional space-time. The defect was modeled as an impenetrable to matter field finite-thickness tube with magnetic flux inside. We implemented the most general form of the Robin boundary condition on the surface of the magnetic tube, which enables a fully general analysis of the problem.
We have found that in flat spacetime, the total vacuum energy generated by a magnetic topological defect depends on the curvature $\xi$, except for special cases corresponding to the Dirichlet and Neumann boundary conditions. By contrast, when Robin's general boundary conditions are imposed, the induced vacuum energy acquires an explicit dependence on the curvature coupling $\xi$, which is significant even in flat space-time.  A detailed study of the dependence of the effect on the boundary condition parameter has been carried out. The obtained results highlight the nontrivial role played by boundary conditions in vacuum polarization phenomena.

\medskip


\end{abstract}

\section{Introduction}

Following the pioneering paper of Casimir \cite{Casimir:1948dh}, it has been understood that external boundaries can lead to nontrivial modifications of the vacuum energy density arising in quantum field theory. 
As a result, Casimir’s original idea stimulated a broad and systematic investigation of quantum vacuum phenomena in the presence of boundaries of various geometrical configurations and composed of different materials. These studies have explored how the shape, size, and physical properties of boundaries influence vacuum fluctuations and associated observable effects, see, e.g., \cite{Elizalde:1995hck,Mostbib,Bordag:2001qi,Klimchitskaya:2023niz}.

In the present paper, we analyze the phenomenon of vacuum polarization  (change of vacuum energy) generated by the presence of a linear magnetic topological defect. The defect is represented by an infinitely long cylindrical finite-thickness tube, impenetrable to the quantized matter field, and carrying a magnetic flux confined within its interior. Such a configuration provides a convenient and physically motivated setting for studying the connections between topology, boundary conditions, and quantum vacuum effects.

Within the framework of second-quantized field theory, the magnetic defect gives rise to a number of nontrivial vacuum phenomena in the region exterior to the tube. In particular, it induces the vacuum energy, generates a vacuum current flowing around the defect, and leads to the emergence of an associated magnetic flux outside the tube, see, e.g.,  {pioneering papers \cite{Serebryanyi:1985blr,Flekkoy:1990pn,Gornicki:1990kq,Bordag1991,Bordag1999}}. Since these phenomena originate from the combined action of the boundary condition imposed on the tube’s surface and the magnetic flux enclosed within the defect, these quantum effects are generally recognized as the Casimir–Bohm–Aharonov effect \cite{Sitenko:1997zf}, emphasizing their relation to both the boundary-induced Casimir effect and topological Aharonov–Bohm-type contributions \cite{Aharonov1959}.

We consider the charged scalar matter field ($\psi$) that interacts with the space-time curvature and is described by the Lagrangian density
\begin{equation}\label{0}
\mathcal{L}=({\mbox{$\nabla$}}_\mu\psi)^*({\mbox{$\nabla$}}^\mu\psi)-(m^2+\xi R)\psi^*\psi,
\end{equation}
where $\xi$ is the scalar–curvature coupling constant (curvature coupling), i.e., the parameter describing the scalar field’s interaction with the curvature of space-time $R$, $\nabla_\mu\psi\equiv\left(\partial_\mu-ie A_\mu\right)\psi$ is the covariant derivative, which includes the vector field  $A_\mu$ associated with a $U(1)$
gauge symmetry.
The energy–momentum tensor (EMT) is obtained by varying the action
$S=\int d^4x\,\sqrt{-g}\, \mathcal{L}$ with respect to the metric tensor 
\cite{chernikov1968quantum,Callan:1970ze,Birrell:1982ix}
\begin{equation}\label{a1}
   T^{\mu\nu}=-\frac{2}{\sqrt{-g}}\frac{\delta S}{\delta g^{\mu\nu}}=T^{\mu\nu}_{can}+2\xi\left(g^{\mu\nu}\Box-\nabla^\mu\nabla^\nu-R^{\mu\nu}\right)
   \psi^*\psi\,,
\end{equation}
where 
\begin{equation}\label{a2}
T^{\mu\nu}_{can}=\nabla^\mu\psi^*\nabla^\nu\psi+
 \nabla^\nu\psi^* \nabla^\mu\psi-g^{\mu\nu}\mathcal{L}
\end{equation}
is the canonical energy-momentum tensor. 
It is evident that the dependence of the scalar field’s EMT on the curvature coupling $\xi$ remains when the Ricci tensor $R^{\mu\nu}=0$, and also in flat space, where the Riemann tensor $R^{\alpha\beta\gamma\delta}=0$. 
 This clearly follows from the fact that the terms arising from varying $\delta\left(\sqrt{-g}R\psi^*\psi\right)$ with respect to the metric tensor after integration by parts lead to terms of the form $\nabla_\mu\nabla_\nu\left(\psi^*\psi\right), \ \Box\,\psi^*\psi$. These terms do not contain a curvature scalar, so they do not vanish with it.

It should be noted that originally, the curvature-coupling term was introduced into the equation of motion for a scalar field with the aim of ensuring conformal invariance of the corresponding field theory. In $(3+1)$-dimensional space-time, this symmetry is realized only for a particular value of the curvature coupling 
$\xi = 1/6$. 
The presence of conformal symmetry has important physical consequences for the properties of the energy–momentum tensor. Specifically, this implies that the trace of the energy–momentum tensor vanishes in the massless theory. As a direct consequence of this, divergences in the Casimir energy associated with flat boundaries are absent, highlighting the regularizing role played by conformal invariance in vacuum energy calculations \cite{Mostbib,Bordag:2001qi}. An analogous situation arises in the case of the electromagnetic field, where conformal invariance also plays an important conceptual and technical role. Nevertheless, there are no fundamental reasons to exclude or disregard alternative values of the curvature coupling beyond the conformal one.

The curvature coupling $\xi$ can be directly interpreted as the sensitivity of the scalar field to curvature $R$ \cite{Calmet2018,Hrycyna2020}. It is interesting to note that the coefficient of R in the effective action of quantized fields after regularization plays the role of the "elasticity" of space in the well-known Sakharov hypothesis, according to which gravitational interaction is not fundamental, but is the effective "elasticity" of space-time arising from quantum fluctuations of matter \cite{Visser2002}.

For a scalar field, a nonvanishing value of the curavure coupling $\xi$ leads to a modification of the effective mass of particles, thereby influencing a wide range of physical processes. In particular, it affects mechanisms of particle creation and decay, plays a significant role during the inflationary stage of the Universe’s evolution \cite{Kaiser:2015usz}, and has important implications for dark matter phenomenology and the formation of large-scale cosmic structures \cite{Sankharva:2021spi}. Furthermore, the value of $\xi$ impacts the stability of the electroweak vacuum \cite{Branchina:2019tyy}, among other phenomena. Despite these numerous theoretical and phenomenological motivations, it remains an open question what exact value of the curvature coupling occurs in nature, and it continues to be the subject of ongoing research.

The present paper focuses on changes in the vacuum energy of a charged massive scalar field $\psi$ in flat space-time in the presence of a magnetic tube of radius $r_0$, which is impenetrable to the matter field \cite{Sitenko:2022gha}. Our analysis focuses on how the induced vacuum energy depends on the curvature coupling. On the boundary of the tube, we adopt the most general Robin-type boundary condition for the field
\begin{equation}\label{Robin}
    (\cos \theta\, \psi + \sin \theta\, r \partial_r \psi)|_{r_0} =0,
\end{equation}
where $\theta$ serves as a parameter of the boundary condition and varies within the interval $-\pi/2 \leq \theta < \pi/2$.
The limiting values $\theta = 0$ and $\theta = -\pi/2$ reproduce the Dirichlet and Neumann boundary conditions, respectively.
The vacuum energy induced in these particular cases has been previously studied in  \cite{Gorkavenko:2009qn,Gorkavenko:2011qg,Gorkavenko:2013rsa,Gorkavenko:2022xtv}.

In \cite{Gorkavenko:2024vuy}, it was demonstrated that in a flat two-dimensional space, the total vacuum energy of a scalar field induced by an impenetrable cylindrical tube characterized by Robin-type boundary conditions and an internal magnetic flux exhibits a dependence on the curvature coupling $\xi$.
In the present paper, we aim to extend this result to the physically relevant case of three-dimensional space.


\section{Induced vacuum energy density. General relations}

In the case of a
static background \cite{Birrell:1982ix,Grib:1980aih,Parker:2009uva}, the operator of the second-quantized charged scalar field takes the form
\begin{equation}\label{a11}
\Psi(x^0,\textbf{x})=\sum\hspace{-1.2em}\int\limits_{\lambda}\frac1{\sqrt{2E_{\lambda}}}
\left[e^{-iE_{\lambda}x^0}\psi_{\lambda}(\textbf{x})\,a_{\lambda}+
  e^{iE_{\lambda}x^0} \psi_\lambda^\ast(\textbf{x})\,b^\dag_{\lambda}\right].
\end{equation}
Here $a^\dag_\lambda$ and $a_\lambda$ ($b^\dag_\lambda$ and
$b_\lambda$) denote the creation and annihilation operators for scalar particles (antiparticle). The label $\lambda$ represents the complete set of quantum numbers that uniquely characterize a given state. 
The quantity  $E_\lambda=E_{-\lambda}>0$ is the corresponding energy of the state. The notation
  $\sum\hspace{-1em}\int\limits_\lambda$ indicates a combined sum over discrete quantum numbers $\lambda$ and integration over continuous ones with an appropriate measure.
Wave functions $\psi_\lambda(\textbf{x})$ satisfy the stationary Klein–Gordon equation
\begin{equation}\label{a12}
 \left\{-{\mbox{\boldmath $\nabla$ }}^2  + (m^2+\xi R)\right\}  \psi_\lambda(\textbf{x})=E^2_\lambda\psi(\textbf{x}),
\end{equation}
where $\mbox{\boldmath $\nabla$ }$ stands for the covariant spatial derivative in the external background field.

The vacuum energy density is defined as the expectation value of the $T^{00}$ component of the energy–momentum tensor in the vacuum state.
In flat space-time, it can be written as
\begin{equation}\label{a14a}
\varepsilon=\varepsilon_{can}+(1/4-\xi)\varepsilon_\xi=\sum\hspace{-1.2em}\int\limits_{\lambda}E_\lambda\psi^*_\lambda(\textbf{x})\,\psi_\lambda(\textbf{x})+(1/4-\xi)\mbox{\boldmath
 $\nabla$}^2
   \sum\hspace{-1.2em}\int\limits_{\lambda}E^{-1}_\lambda\psi^*_\lambda(\textbf{x})\,\psi_\lambda(\textbf{x}),
\end{equation}
 where $\varepsilon_{can}$ denotes the canonical energy density $T^{00}_{can}$, which is independent of $\xi$. 
It should be noted that for $\xi=1/4$, the energy density reduces to its canonical form \cite{Sitenko:2002zp}.

We consider a static background consisting of a cylindrical tube of finite thickness that contains a magnetic flux.
The coordinate system is chosen such that the tube is aligned with the $z$-axis.
The covariant derivative is defined as $\nabla_0=\partial_0$, $\mbox{\boldmath
$\nabla$}=\mbox{\boldmath $\partial$}-{\rm i}  e\, {\bf V}$. The vector potential has a single nonvanishing component outside the tube, 
\begin{equation}\label{4}
V_\varphi=\Phi/2\pi,
\end{equation}
where $\Phi$ denotes the magnetic flux confined within the tube, and $\varphi$ corresponds to the angular variable in the polar coordinate system $(r,\varphi)$ on the plane orthogonal to the tube’s axis.
  
In flat space-time, the function satisfying Eq.\eqref{a12} and obeying the Robin boundary conditions \eqref{Robin} at the surface of an impenetrable tube with radius $r_0$ can be expressed as follows in the region outside the tube,  {see Appendix,}\vspace{-0.5em}
\begin{equation}\label{6}
\psi_{kn{\bf p}}({\bf x})=\frac{e^{{\rm i}n\varphi}}{2\pi}\,e^{{\rm i} p
z}\Omega_{|n- {e}
\Phi/2\pi|}(\theta,kr,kr_0),
\end{equation}
where $\theta$ is the parameter characterizing the boundary condition \eqref{Robin},
\begin{align}
& \Omega_\rho(\theta,u,v)=\sin\mu_\rho(\theta,v) J_{\rho}(u)-\cos\mu_\rho(\theta,v) Y_{\rho}(u),\label{7} \\
    & \sin\mu_\rho(\theta,v)=[\gamma_\rho(\theta,v)]^{-1/2}[\cos\theta\, Y_{\rho}(v)+\sin\theta\, v Y'_{\rho}(v)],\\
  &  \cos\mu_\rho(\theta,v)=[\gamma_\rho(\theta,v)]^{-1/2}[\cos\theta\, J_{\rho}(v)+\sin\theta\, v J'_{\rho}(v)],\\
  &  {\gamma_\rho(\theta,v)=\left[\cos\theta J_{\rho}(v)+\sin\theta\, v J'_{\rho}(v)\right]^2+\left[\cos\theta\, Y_{\rho}(v)+\sin\theta\, v Y'_{\rho}(v)\right]^2.}
\end{align}
 {Here, $\gamma_\rho(\theta,v)$ denotes the factor that ensures the proper normalization of the functions,} $ 0 < k < \infty$, $ -\infty < p < \infty $, and index $n$ takes values from $\mathbb{Z} $, the set of all integers. The symbols $ J_\rho(u) $ and $ Y_\rho(u) $ represent the Bessel functions of the first and second kinds, respectively, of order 
 $ \rho $ \cite{AbramowitzStegun}. The prime notation applied to a function indicates differentiation with respect to its argument.

Unfortunately, the definition \eqref{a14a} exhibits ultraviolet divergences and requires renormalization.  {For the considered setup, where the magnetic field is confined outside the region available to the matter field, the renormalization procedure requires only a single subtraction: removing the contribution corresponding to vanishing magnetic flux \cite{Babansky:1999re}. The induced vacuum energy is defined as the difference between the vacuum energies of an impenetrable tube with and without magnetic flux, evaluated at a fixed value of the tube thickness. The surface divergence associated with the tube boundary (Casimir surface energy), which depends only on the boundary geometry and imposed boundary conditions \cite{Mostbib,Bordag:2001qi,Bordag:2001ta,Vassilevich:2003xt}, is independent of the magnetic flux and therefore cancels in the subtraction. For this reason, when computing the induced vacuum energy, we do not separate the contributions of the bulk and surface divergent terms, see, e.g., \cite{Braganca:2020jci}. Therefore,}
the renormalized vacuum energy density can be written as \cite{Gorkavenko:2013rsa}\vspace{-0.5em}
\begin{multline}\label{c2qq}
\varepsilon_{ren}=\varepsilon_{can}+(1/4-\xi)\varepsilon_\xi=\\= \frac{1}{(2\pi)^{2}} \int\limits_{-\infty}^\infty d 
p\int\limits_0^\infty
  dk\,k\left(\sqrt{ p^2+k^2+m^2}+\frac{1/4-\xi}{\sqrt{p^2+k^2+m^2}}\triangle_r\right)G(\theta,kr,kr_0,\Phi).
\end{multline}
Here  
$\triangle_r = \partial_r^2 + r^{-1} \partial_r$  
represents the radial component of the Laplacian operator acting on the plane perpendicular to the tube.
The function $G$ is introduced as follows
\begin{equation}\label{Gfunction}
    G(\theta,kr,kr_0,\Phi)=S(\theta,kr,kr_0,\Phi)-S(\theta,kr,kr_0,0),
\end{equation}
\begin{equation}\label{a29a}
S(\theta,kr,kr_0,\Phi)=\sum_{n\in\mathbb
 Z}\Omega^2_{|n- {e}
\Phi/2\pi|}(\theta,kr,kr_0).
\end{equation}
A comprehensive discussion of the properties and different representations of the 
$S$ function can be found in \cite{Gorkavenko:2024vuy}.

It is important to emphasize that because the summation in \eqref{a29a} extends over an infinite range, the function $S$, and consequently the function $G$, along with all related induced physical quantities, depend solely on the fractional part of the magnetic flux contained within the tube. This fractional magnetic flux is defined as
\begin{equation}\label{a29a1}
F=\frac{e\Phi}{2\pi}-\left[\!\left[\frac{e\Phi}{2\pi}\right]\!\right],\quad(0\leq F < 1),
\end{equation}
where $[[u]]$ denotes the integer part of the quantity $u$, i.e., the greatest integer less than or equal to $u$.

Furthermore, the dependence of all physical quantities on $F$
exhibits a symmetry under the transformation  $F\rightarrow 1-F$. This symmetry reflects the topological nature of the effect and is a consequence of the fact that the matter field does not penetrate into the region of space where the magnetic field is present. It therefore represents a clear manifestation of the Aharonov–Bohm phenomenon, emphasizing that only the fractional part of the magnetic flux has physical significance in this setting.

\section{Induced vacuum energy. Dependence on the curvature\\ coupling}

Since the induced vacuum energy density \eqref{c2qq} is a function solely of the radial distance in the plane orthogonal to the cylindrical tube, it exhibits no dependence on the longitudinal coordinate. As a consequence, when one attempts to compute the total induced vacuum energy in $(3+1)$-dimensional space-time by integrating this density over the entire spatial volume, the integration along the $z$-direction (perpendicular to the transverse plane of the tube) leads to a divergence. 

For this reason, from a physical point of view, it is more appropriate and meaningful to restrict the analysis to the induced vacuum energy defined per unit length of the tube, or equivalently, to consider the energy integrated only over the plane orthogonal to the tube. This approach yields a finite and well-defined quantity that captures the physically relevant vacuum polarization effects associated with the presence of the magnetic defect.

In this paper, we will focus only on the contribution to the induced vacuum energy \eqref{c2qq} in the plane orthogonal to the tube that is proportional to $(1/4-\xi)$.   {By exchanging the order of integration over $r$ and $p$, this contribution to the induced vacuum energy can be rewritten in terms of the corresponding induced vacuum energy in two-dimensional space-time; for details, see \cite{Gorkavenko:2013rsa,Gorkavenko:2025tml}:}
\begin{equation}\label{4s5}
E^{(3+1)}_{\xi}=\frac{1}{2\pi}\int\limits_{r_0}^\infty dr \,r \int\limits_{-\infty}^\infty d \,
p\int\limits_0^\infty
  dk\,k\,\frac{\Delta_r G(\theta,kr,kr_0,F)}{\sqrt{ p^2+k^2+m^2}}
  =\frac{1}{(r_0)^{2}}\frac{1}{\pi}\int\limits_{mr_0}^\infty
dv\, \frac{v^2 \,\mathcal{D}_\xi(\theta,v,F)}{\sqrt{v^2-(mr_0)^2}},
\end{equation}
where $\mathcal{D}_\xi$ determines the induced vacuum energy in two-dimensional space-time
\begin{equation}\label{dop3}
E^{(2+1)}_{\xi}= m \mathcal{D}_{\xi}(\theta,mr_0,F),
\end{equation}
\begin{equation}\label{dddop4}
\mathcal{D}_{\xi}(\theta,y,F)=\int\limits_{y}^\infty dx\,x \Delta_x \int\limits_0^\infty
dz\, z\frac{G(\theta,z,z y/x,F)}{x\sqrt{z^2+x^2}}.
\end{equation}

 \begin{figure}[t!]
    \centering
    \includegraphics[width=\textwidth]{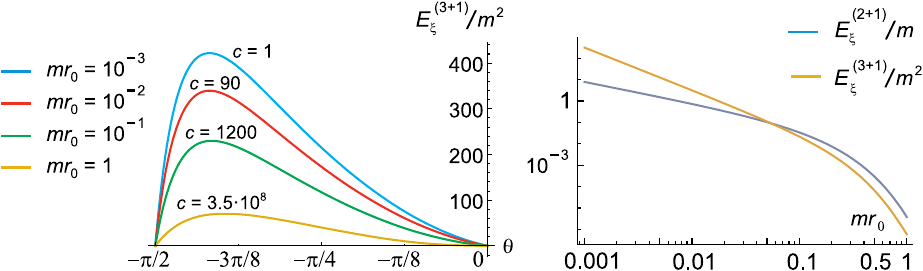}
    \caption{Left panel: the induced dimensionless vacuum energy $E_\xi^{3+1}/m^{2}$ is plotted as a function of the parameter $\theta$, which characterizes the Robin boundary conditions on the surface of the impenetrable magnetic tube, for various tube thicknesses. For clarity, the curves are multiplied by the coefficient $c$. Right panel: comparison of the induced dimensionless vacuum energy in $(3+1)$- and $(2+1)$-dimensional space-times as functions of the dimensionless tube thickness $mr_0$ with the boundary condition parameter $\theta=-1$.
    For both panels, we restrict ourselves to a half-integer magnetic flux $F = 1/2$.}
    \label{Fig1}
\end{figure}

Using the method obtained in \cite{Gorkavenko:2024vuy} for the vacuum energy $E^{(2+1)}_\xi$ in $(2+1)$-dimensional flat space-time induced by a magnetic tube of thickness $mr_0=1/100$, we numerically compute $E^{(2+1)}_\xi$ for other tube thicknesses. In our analysis, we fix the magnetic flux inside the tube at the half-integer value $F=1/2$, for which the induced vacuum energy is known to be maximal, as established analytically in the case of a singular magnetic vortex \cite{Sitenko:2002zp,Serebryanyi:1985blr}. In the case of a magnetic tube of finite radius, we expect the induced vacuum energy to be a more complicated function of the magnetic flux, but to remain proportional to $\sin(F\pi)$, as in the case of a singular magnetic vortex. For simplicity, we also restrict ourselves to considering only not positive values of the parameter characterizing the boundary condition $-\pi/2\leq\theta\leq0$, since for positive values of parameter $\theta$ the calculations would have to be more involved, taking into account the emergence of bound states in the solutions of \eqref{a12}, see \cite{Sitenko:2022gha}.

We computed the induced vacuum energy  numerically 
 in $(3+1)$-dimensional flat space-time and compared the results with those obtained in $(2+1)$-dimensional space-time, as illustrated in Fig.\ref{Fig1}.  
One can see that the induced dimensionless vacuum energy in three-dimensional flat space-time, $E^{(3+1)}_\xi/m^{2}$, induced by the presence of the magnetic topological defect is independent of the curvature coupling $\xi$ only in the limiting cases of Dirichlet ($\theta=0$) and Neumann ($\theta=-\pi/2$) boundary conditions,  {where $E^{(3+1)}_\xi/m^{2}=0$. For intermediate values, $-\pi/2<\theta<0$, the induced vacuum energy $E^{(3+1)}_\xi/m^{2}$ is nonzero.}
One can also see that $E^{(3+1)}_\xi/m^{2}$ rapidly decreases with increasing tube thickness $mr_0$, see left panel.  
Furthermore, a comparison of the induced vacuum energy in three- and two-dimensional space-time for the partial case of boundary condition parameter $\theta=-1$ shows that for thin magnetic tubes ($mr_0\lesssim 0.05$), the dimensionless induced vacuum energy is larger in $(3+1)$-dimensional space–time. In contrast, for thick magnetic tubes ($mr_0\gtrsim 0.05$), the dimensionless induced vacuum energy becomes larger in $(2+1)$-dimensional space–time, as shown in the right panel.

\section{Dependence on boundary condition parameter}

In the previous section, we derived the dependence of the induced vacuum energy $E^{(3+1)}_\xi$ on the boundary-condition parameter for various magnetic tube thicknesses, see Fig.\ref{Fig1}. We now focus on a detailed investigation of the 
$\theta$-dependence of more subtle features of the vacuum effect.

At first, let us discuss the asymptotic behavior of the induced vacuum energy $E^{(3+1)}_\xi$ for the case of infinitely thin and sufficiently thick tubes. Please note that since quantum effects depend on the dimensionless thickness of the tube $mr_0$, there are actually two possible scenarios: either the tube radius $r_0$ or the scalar field mass $m$ becomes infinitely large or small.

The asymptotic relation for the case of $mr_0\ll 1$ can be obtained from \eqref{4s5}
\begin{equation}\label{4s5new}
E^{(3+1)}_{\xi}
  =\frac{C(\theta,F)}{(r_0)^{2}},\quad C(\theta,F)=\frac{1}{\pi}\int\limits_{0}^\infty
dv\, v \,\mathcal{D}_\xi(\theta,v,F).
\end{equation}
It should be noted that the asymptotic dependence of the induced energy $\sim (r_0)^{-2}$ 
is fully consistent with the results obtained in the approach of the singular magnetic vortex, see, e.g., \cite{Babansky:1999re,Sitenko:2002zp,Serebryanyi:1985blr}.  As one can see, relation \eqref{4s5new} does not depend on mass and contains only one dimensional parameter $r_0$. So, relation $E^{(3+1)}_{\xi}\sim (r_0)^{-2}$ can be obtained only from dimensional analysis\footnote{We would like to remind you that $E^{(3+1)}_{\xi}$ is not the full induced vacuum energy, but the corresponding energy density integrated only over the coordinates in the plane perpendicular to the tube.}.

\begin{figure}[t!]
    \centering
    \includegraphics[width=1\textwidth]{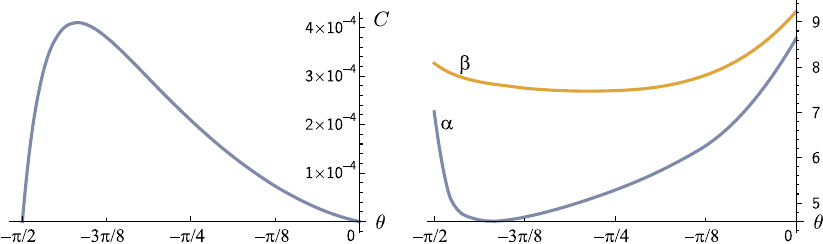}
    \caption{Left panel: the dependence of the coefficient $C(\theta,F)$ in the asymptotic relation at $mr_0\ll 1$ \eqref{4s5new} on boundary condition parameter $\theta$. Right panel: the dependence of the coefficients $\alpha(\theta,F)$ and $\beta(\theta,F)$  in the asymptotic relations at $mr_0\gg 1$ \eqref{gg2}, \eqref{gg} on boundary condition parameter $\theta$. We restrict ourselves to a half-integer magnetic flux $F = 1/2$.}
    \label{Fig2}
\end{figure}

The dependence of the coefficient $C(\theta,F)$ in \eqref{4s5new} on boundary condition parameter $\theta$ at half-integer magnetic flux inside the tube $F=1/2$ is presented in Fig.\ref{Fig2}. As one can see, this dependence qualitatively reproduces behavior of the induced vacuum energy $E^{(3+1)}_{\xi}$ as a function of parameter $\theta$, see Fig.\ref{Fig1}.

The asymptotic relation for the case of $mr_0\gg 1$ follows from \eqref{4s5} by employing the asymptotic expression for the function $\mathcal{D}_\xi$, i.e., for the induced vacuum energy in 
$(2+1)$-dimensional space, see \eqref{dop3}. 
Numerical analysis indicates that this function can be well approximated at large $mr_0$
as
\begin{equation}\label{gg2}
    \mathcal{D}^{as}_\xi=\frac{e^{-\alpha -\beta \,mr_0}}{mr_0}, \quad mr_0 \gg 1,
\end{equation}
where $\alpha$, $\beta$ are positive coefficients depending on the parameters $\theta$ and $F$.
Substituting this asymptotic form into \eqref{4s5}, the integration can be performed explicitly, yielding
\begin{equation}\label{gg}
 E^{(3+1)}_{\xi}=  m^{2} \frac{e^{-\alpha}}{ \pi \, mr_0} K_1(\beta\,  mr_0)  {\approx  \frac{m^{2}}{\sqrt{2 \pi \beta} } \frac{e^{-\alpha- \beta\, mr_0}}{( mr_0)^{{3}/{2}}},} \quad x_0\gg 1,
\end{equation}
where $K_\mu(u)$ denotes the Macdonald function. 

The dependence of the coefficients $\alpha(\theta,F)$ and $\beta$ in \eqref{4s5new} on the boundary condition parameter $\theta$ at half-integer magnetic flux inside the tube $F=1/2$ is presented in Fig.\ref{Fig2}. One can observe the expected correlation: if the induced vacuum energy at a given $\theta$ is larger at intermediate values of the parameter $mr_0$, then it also decreases more slowly as $mr_0$ increases.

Finally, we became interested in the question of the tube thickness $mr_0$
 below which the dimensionless induced vacuum energy in $(3+1)$-dimensional space-time becomes larger than that in $(2+1)$-dimensional space-time. This behavior is illustrated in Fig.\ref{Fig1}, where the situation is shown for a particular value of the parameter characterizing the boundary condition, $\theta=-1$. We investigated the value of the tube thickness $mr_0$ at which the dimensionless induced vacuum energy in 
$(3+1)$-dimensional space-time equals the corresponding dimensionless induced energy in 
$(2+1)$-dimensional space-time as a function of the boundary-condition parameter, see Fig.\ref{Fig3}. Somewhat unexpectedly, our analysis shows that the corresponding value of the magnetic tube thickness is practically independent of 
$\theta$ and is approximately $mr_0\approx 0.05$.

\section{Conclusions} 

We studied vacuum polarization (change of vacuum energy) of a charged massive scalar field induced by a magnetic topological defect in flat $(3+1)$-dimensional spacetime.  The defect was modeled as an impenetrable to matter field finite-thickness tube with magnetic flux inside. On the boundary of the tube, we adopt the
most general Robin-type boundary condition for the field, allowing for an analysis of vacuum effects in a fully general setting. Our attention was focused on studying the dependence of the induced vacuum energy in flat space-time on the curvature coupling $\xi$, which governs the interaction between the scalar field and the space-time curvature.

\begin{figure}[t!]
    \centering
    \includegraphics{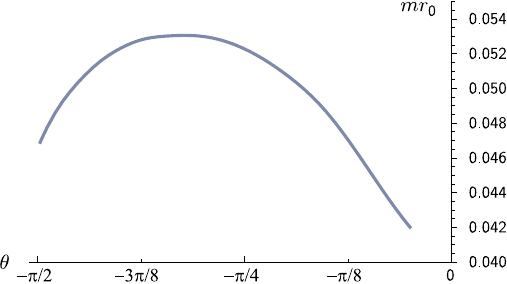}
    \caption{The dependence of the tube thickness $mr_0$, at which the dimensionless induced vacuum energy  {$E^{(3+1)}_{\xi}/m^2$} in $(3+1)$-dimensional space-time equals the corresponding dimensionless induced vacuum energy  {$E^{(2+1)}_{\xi}/m$} in $(2+1)$-dimensional space-time, on the boundary condition parameter $\theta$ for $F=1/2$.}
    \label{Fig3}
\end{figure}

We find that, in flat space-time, the  induced vacuum energy within the plane perpendicular to the
tube can be written as  $E^{(3+1)}=E^{(3+1)}_{can}+(1/4-\xi)E^{(3+1)}_\xi$ and does not depend on the curvature coupling 
$\xi$ only in the special cases of Dirichlet and Neumann boundary conditions, i.e., when the induced vacuum energy $E^{(3+1)}_\xi$ vanishes. 
In contrast, when general Robin boundary conditions are imposed, the induced vacuum energy exhibits an explicit dependence on the curvature coupling $\xi$ and demonstrates a nontrivial sensitivity to the boundary condition parameter  $\theta$.  Our conclusion is that, in principle, precise experimental observations of vacuum polarization in flat space-time may serve as an independent approach to investigate the curvature coupling parameter $\xi$.

For values of $\theta$ in the range $-\pi/2 \leq \theta \leq 0$, we numerically evaluated $E^{(3+1)}_\xi$, see the left panel of Fig.\ref{Fig1}. As can be observed from our analysis, the magnitude of the induced vacuum energy increases significantly as the thickness of the magnetic tube $mr_0$ decreases. However, for magnetic tubes of large thickness $mr_0\gtrsim 1$, the induced vacuum energy decreases very rapidly. The dependence of $E^{(3+1)}_\xi$ on the boundary condition parameter is nontrivial: it is a smooth function with a single maximum. As the thickness of the tube decreases, the boundary condition parameter  $\theta$ at which the induced energy attains its maximum shifts gradually toward lower values, approaching $\theta = -1.313$ in the limit of an infinitely thin tube, see the left panel of Fig.\ref{Fig2}.

We carried out a detailed analysis of the asymptotic behavior of the induced vacuum energy $E^{(3+1)}_\xi$ in the limiting cases of an ultra-thin ($mr_0\ll 1$) and significantly thick ($mr_0\gg1$) magnetic tube, see the corresponding relations \eqref{4s5new} and \eqref{gg}. Additionally, we examined the dependence of the asymptotic coefficients on the boundary condition parameter $\theta$, see Fig.\ref{Fig2}.

Finally, we became interested in the question of the tube thickness $mr_0$
 below which the dimensionless induced vacuum energy in $(3+1)$-dimensional space-time becomes larger than that in $(2+1)$-dimensional space-time, see right panel of Fig.\ref{Fig1} for a particular value of the boundary condition parameter $\theta=-1$. We studied the tube thickness $mr_0$, at which the induced dimensionless vacuum energy in $(3+1)$-dimensional spacetime coincides with that in $(2+1)$-dimensional spacetime, depends on the boundary condition parameter $\theta$, see Fig.~\ref{Fig3}. Remarkably, this characteristic thickness turns out to be nearly independent of $\theta$ and is approximately $mr_0 \approx 0.05$.

 It would be interesting to further investigate the vacuum energy induced by a magnetic tube with Robin boundary conditions and a positive boundary parameter. In this regime, the Klein–Gordon equation may contain bound state solutions, which can make a significant contribution to the induced vacuum energy. This could yield interesting and nontrivial results similar to those discussed in \cite{Sitenko:2022gha},  {and in \cite{Bordag:2025ets} for the Casimir effect, where this was done for the first time.}

\setcounter{equation}{0}
\renewcommand{\theequation}{A.\arabic{equation}}
\section*{Appendix: Solutions to the  Klein–Gordon equation}

 The Klein-Gordon equation \eqref{a12} in the case of flat space-time takes the form
\begin{equation}
\{-(\mbox{\boldmath $\nabla$ }^{2}+m^{2})\}\Psi_{\lambda}(\textbf{x})=E_{\lambda}^{2}\Psi_{\lambda}(\textbf{x}).
\end{equation}
In the external field background given by Eq.\eqref{4}, using cylindrical coordinates, we obtain
\begin{equation}
\label{A2}
\left\{-\left(\frac{\partial^{2}}{\partial r^{2}} + \frac{1}{r^{2}}\left(\frac{\partial}{\partial\varphi} - i \frac{e \Phi}{2\pi}\right)^{2} + \frac{1}{r}\frac{\partial}{\partial r} + \frac{\partial^{2}}{\partial z^{2}}\right) + m^{2}\right\}\Psi_{\lambda}(\boldsymbol{x}) = E_{\lambda}^{2}\Psi_{\lambda}(\boldsymbol{x}).
\end{equation}

We will seek a solution in the form
\begin{equation}
\label{A3}
\Psi_{\lambda}(\boldsymbol{x}) = \beta f(kr) e^{i n\varphi} e^{ip z}. 
\end{equation}
Substituting this ansatz into Eq. \eqref{A2}, and using the relation $E^{2} = m^{2} + p^{2} + k^{2}$, we obtain the radial equation
\begin{equation}
\frac{\partial^{2}f}{\partial r^{2}} + \frac{1}{r}\frac{\partial f}{\partial r} + \left(k^{2} - \frac{(n - \frac{e \Phi}{2\pi})^{2}}{r^2}\right)f = 0.
\end{equation}
Introducing the dimensionless variable $\alpha = kr$, this equation takes the form of the Bessel equation,
\begin{equation}
\alpha^{2}\frac{\partial^{2}f}{\partial\alpha^{2}} + \alpha\frac{\partial f}{\partial\alpha} + \left(\alpha^{2} - \left(n - \frac{e\Phi}{2\pi}\right)^{2}\right)f = 0, 
\end{equation}
whose general solution is
\begin{equation}\label{A6}
f(kr) = A J_{\tau}(kr) + B Y_{\tau}(kr),
\end{equation}
where $J_\tau(u)$ and $Y_\tau(u)$ represent the Bessel functions of the first and second kinds, respectively, of order 
 $ \tau $ \cite{AbramowitzStegun}; $\tau = |n - \frac{e\Phi}{2\pi}|$. 
 
By imposing the Robin boundary condition \eqref{Robin}, we find that parameters $A$ and $B$ in \eqref{A6} can be presented in the form
\begin{equation}
A=\cos\theta\, Y_{\tau}(kr_0)+\sin\theta\, (kr_0) Y'_{\tau}(kr_0),
  \quad  B=\cos\theta\, J_{\tau}(kr_0)+\sin\theta\, (kr_0) J'_{\tau}(kr_0).
\end{equation}

The general normalization factor $\beta$ in \eqref{A3} can be found from the normalization condition, which, for our case of flat space-time outside the tube with an impenetrable boundary and a static background, takes the form
\begin{equation}\label{norma}
\int\limits_{r>r_0} \Psi_{knp}^{*}(\boldsymbol{x}) \Psi_{k'n'p'}(\boldsymbol{x}) d\boldsymbol{x} = \delta_{nn'} \delta(p-p') \frac{\delta(k - k')}{k}.
\end{equation}
The factor $k^{-1}$
on the right-hand side is related to the fact that $k$
is the radial component of the momentum in the cylindrical coordinate system, reflecting the measure in momentum space.
The choice of normalization condition in the form \eqref{norma} is consistent with the expression for the second-quantized charged scalar  field operator \eqref{a11}, in which the factor $1/\sqrt{2E_{\lambda}}$ is explicitly singled out.

Thus, we obtain
\begin{equation}
    \beta=\frac{1}{2\pi}\frac{1}{\sqrt{A^2+B^2}}.
\end{equation}

Finally, by introducing the notation 
$\sin\mu=A/\sqrt{A^2+B^2}$ and $\cos\mu=B/\sqrt{A^2+B^2}$, we obtain Exp.\eqref{7}.

\section*{Acknowledgments}
V.G. would like to warmly thank the organizers of the “BGL2025: The 13th Bolyai–Gauss–Loba\-chev\-sky Conference on Non-Euclidean Geometry and Its Applications” for their kind hospitality and for creating such a welcoming and inspiring atmosphere throughout the meeting.
The work of V.G. was partially supported by the project 'High-energy processes in plasma: acceleration of cosmic rays and their contribution to space weather' of the Ministry of Education and Science of Ukraine (25BF051-04, 0125U002256).

\bibliography{bibliography}
\bibliographystyle{JHEP}

\end{document}